\documentclass[aps,prb,reprint,floatfix,amsmath,amssymb]{revtex4-2}
\usepackage[T1]{fontenc}
\usepackage{bm}
\usepackage{amsthm,mathtools}
\usepackage{microtype,booktabs}
\usepackage{tikz-cd}
\usepackage[bookmarksnumbered=true,colorlinks=true,allcolors=blue]{hyperref}
\hypersetup{pdftitle={Computing free-fermion topology in real space},pdfauthor={Tian Yuan, Yang Qi}}
\usepackage[nameinlink,noabbrev]{cleveref}
\crefname{equation}{Eq.}{Eqs.}
\Crefname{equation}{Equation}{Equations}
\crefname{section}{Sec.}{Secs.}
\Crefname{section}{Section}{Sections}
\crefname{appendix}{Appendix}{Appendices}
\Crefname{appendix}{Appendix}{Appendices}
\crefname{figure}{Fig.}{Figs.}
\Crefname{figure}{Figure}{Figures}
\crefname{table}{Table}{Tables}
\Crefname{table}{Table}{Tables}
\crefname{subsection}{Sec.}{Secs.}
\Crefname{subsection}{Section}{Sections}
\theoremstyle{plain}
\newtheorem{prop}{Proposition}[section]

\theoremstyle{definition}

\newtheorem{conj}[prop]{Conjecture}
\theoremstyle{remark}
\newtheorem{rk}[prop]{Remark}
\crefname{prop}{proposition}{propositions}
\crefname{conj}{Conjecture}{Conjectures}
\Crefname{conj}{Conjecture}{Conjectures}
\crefname{rk}{remark}{remarks}

\newcommand{\symalg}{A_{\mathrm{sym}}}
\newcommand{\realcls}{L}
\newcommand{\clifford}{\mathrm{Cl}}
\newcommand{\Lat}{\mathsf{Lat}}
\newcommand{\Sph}{\mathsf{Sph}}
\newcommand{\Disc}{\mathsf{Disc}}
\DeclareMathOperator{\sgn}{sgn}
\DeclareMathOperator{\spec}{spec}

\DeclareMathOperator{\coker}{coker}

\DeclareMathOperator{\ind}{ind}
\newcommand{\R}{\mathbb{R}}
\newcommand{\C}{\mathbb{C}}
\newcommand{\Z}{\mathbb{Z}}

\begin{document}

\title{Computing free-fermion topology in real space}
\author{Tian Yuan}
\email{tyuan19@fudan.edu.cn}
\author{Yang Qi}
\email{qiyang@fudan.edu.cn}
\affiliation{State Key Laboratory of Surface Physics and Department of Physics, Fudan University, Shanghai 200433, China}
\date{November 26, 2026}

\begin{abstract}
We present a construction for computing topological classes of gapped tight-binding free-fermion Hamiltonians without assuming translation symmetry. We formulate Kitaev's real-space classification proposal as a conjecture for general symmetry data with point-group symmetries, antiunitary symmetries, antisymmetries that anticommute with the Hamiltonian, and projective phases. From this framework, we obtain an algorithm that reduces the input Hamiltonian to a finite-size zero-dimensional system and computes its class in the corresponding $K$-group. This gives a classifying-space interpretation of the spectral localizer. Our GAP package FreeFermionicSPT computes the classification group and the element represented by the finite-size Hamiltonian.
\end{abstract}
\maketitle

\section{Introduction}\label{sec-intro}

Topological insulators and superconductors link the topology of a gapped bulk Hamiltonian to quantized responses and protected boundary states~\cite{thoulessQuantizedHallConductance1982,kane2TopologicalOrder2005,kitaevUnpairedMajoranaFermions2001,schindlerHigherorderTopologyBismuth2018}. The tenfold way and its crystalline extensions organize free-fermion phases by dimension and symmetry~\cite{schnyderClassificationTopologicalInsulators2008,kitaevPeriodicTableTopological2009,ryu2010tenfold,fuTopologicalCrystallineInsulators2011,cornfeldClassificationCrystallineTopological2019,okumaTopologicalClassificationNonmagnetic2019}. Applying these classifications to a concrete system requires invariants that can be evaluated from its Hamiltonian.

In momentum space, Bloch wave functions and symmetry representations lead to practical formulas for topological invariants~\cite{thoulessQuantizedHallConductance1982,yu2011equivalent,fuTopologicalInsulatorsInversion2007,poSymmetrybasedIndicatorsBand2017,bradlynTopologicalQuantumChemistry2017}. Symmetry indicators simplify the calculation, although they do not distinguish every topological phase~\cite{poSymmetrybasedIndicatorsBand2017}.

In this work, an algorithm is proposed to compute the topological indices for a given gapped tight-binding Hamiltonian in real space. Point-group symmetries may be present, together with antiunitary operations, symmetries that anticommute with the Hamiltonian, and projective phases. Our construction treats these data together and assigns a class to the given Hamiltonian. Our calculation also applies in disordered and nonperiodic systems without translation symmetry. Several other approaches also compute topological invariants directly in real space~\cite{bellissard1994noncommutative,loring2010disordered,bianco2011mapping,fulga2012scattering,velury2025markers}.

Our starting point is Kitaev's proposal for local definitions of gapped Hamiltonians~\cite{alexeikitaevLocalDefinitionsGapped}. We formulate this proposal for general symmetry data in \cref{sec-spaces}. The resulting conjecture asserts an equivalence between the real-space classifying space and a zero-dimensional one with extra symmetry constraints. The comparison maps yield an algorithm for computing the class of a given Hamiltonian in the corresponding $K$-group (\cref{sec:algorithm}).

Starting from an $n$-dimensional Hamiltonian $H$, we first enlarge the internal space by tensoring an auxiliary Hilbert space $S$ at each site, taking $H$ to $\Gamma\otimes H$, where $\Gamma$ is the grading on $S$. We then construct a finite-size zero-dimensional Hamiltonian that spatially interpolates between $\Gamma\otimes H$ at the origin and topologically trivial Hamiltonians in regions $x\geq R$. It inherits the physical symmetries and anticommutes with $n$ additional Clifford generators. In the usual periodic table of the tenfold way, these constraints implement the familiar correspondence between phases in symmetry class $a$ in $n$ dimensions and phases in symmetry class $a-n$ in zero dimensions~\cite{kitaevPeriodicTableTopological2009}, formally known as the $(1,1)$-periodicity. With spatial symmetries, the shift depends on how the group acts on space, and we use a generalization of $(1,1)$-Bott periodicity.

Concretely, let $V$ be an $n$-dimensional Euclidean space and $\Lambda\subset V$ the lattice. We write Euclidean vectors in bold. Start from a quadratic Hamiltonian in second quantization,
\begin{equation}
 \widehat H=\sum_{\bm{x},\bm{y}\in\Lambda}\hat\psi_{\bm{x}}^\dagger H_{\bm{x}\bm{y}}\hat\psi_{\bm{y}}.
 \label{eq-intro-second-quantized}
\end{equation}
Here $\hat\psi_{\bm{x}}$ collects the fermion annihilation operators at site $\bm{x}$, and $H_{\bm{x}\bm{y}}$ contains the hopping and onsite matrix elements. For BdG Hamiltonians, replace $\hat\psi_{\bm{x}}$ by a Nambu spinor of annihilation and creation operators. The input also specifies an orthogonal action $\bm{x}\mapsto g\bm{x}=\rho_V(g)\bm{x}$ of a finite group $G$ on $V$, preserving $\Lambda$, and internal single-particle operators $U_g$ satisfying
\begin{equation}
 U_gH_{\bm{x}\bm{y}}U_g^{-1}=(-1)^{h(g)}H_{g\bm{x},g\bm{y}},
 \qquad U_gU_k=\omega(g,k)U_{gk}.
 \label{eq-intro-symmetry-operators}
\end{equation}
The operators $U_g$ are unitary for $s(g)=0$ and antiunitary for $s(g)=1$. The value $h(g)=0$ or $1$ distinguishes commuting and anticommuting operations, and $\omega(g,k)$ records their projective phases. These operators include spatial symmetries, time reversal, and, when present, particle-hole and chiral constraints.

The finite-size zero-dimensional Hamiltonian is constructed in a finite sphereical subspace $|\bm{x}|\leq R$ for $\bm{x}\in V$, on an enlarged Hilbert space.
We tensor the internal space at each site with the $2^n$-dimensional auxiliary space $S=\Lambda^*V$, the real exterior algebra of $V$, carrying matrices $\Gamma,c_j,q_j$ for $j=1,\ldots,n$, where $n=\mathrm{dim}(V)$. The matrices $\Gamma,c_j,iq_j$ are Hermitian, square to $I$, and mutually anticommute. We use them to construct the following Hamiltonian $\hat H_{0\mathrm{d}}=\sum_{\bm{x}\bm{y}}\hat\psi_{\bm{x}}^\dagger m_{\bm{x}\bm{y}}\hat\psi_{\bm{x}}$, where the matirx $m$ is given by
\begin{equation}
 m_{\bm{x}\bm{y}}=\sqrt{v_0(\bm{x})v_0(\bm{y})}\,(\Gamma\otimes H_{\bm{x}\bm{y}})
       +\delta_{\bm{x}\bm{y}}\sum_{j=1}^n v_j(\bm{x})(c_j\otimes I).
 \label{eq-intro-finite-matrix}
\end{equation}
For $|\bm{x}|\le R$, the scalar $v_0$ and the vector $\bm{v}$ with components $v_j$ are given by
\begin{equation}\label{eq:v-def}
 \begin{aligned}
 v_0(\bm{x})&=\cos\!\left(\frac{\pi|\bm{x}|}{2R}\right),\\
 \bm{v}(\bm{x})&=\begin{cases}
   \displaystyle\sin\!\left(\frac{\pi|\bm{x}|}{2R}\right)\frac{\bm{x}}{|\bm{x}|}, & \bm{x}\ne\bm{0},\\[4pt]
   \bm{0}, & \bm{x}=\bm{0}.
 \end{cases}
 \end{aligned}
\end{equation}
The pair $(v_0,\bm{v})$ is a unit vector, with $v_0^2+|\bm{v}|^2=1$. At and beyond radius $R$ we take $v_0=0$ and $\bm{v}(\bm{x})=\bm{x}/|\bm{x}|$. Only decoupled onsite blocks with eigenvalues $\pm1$ remain there, so the calculation uses just the sites with $|\bm{x}|<R$.
By construction, the matrix $m$ anticommutes with $q_i$.
Hence, $q_i$ can be viewed as artificial symmetry generators that changes the symmetry class.

The physical symmetries, including the spatial symmetries, also acts on this extended Hamiltonian.
Write a single-particle state as $|\xi\rangle=\sum_{\bm{x}\in\Lambda_R}|\bm{x}\rangle\otimes|\xi_{\bm{x}}\rangle$, where $|\xi_{\bm{x}}\rangle$ is its internal amplitude at site $\bm{x}$. The physical symmetries act by
\begin{equation}
 \rho^+(g)|\xi\rangle
 =\sum_{\bm{x}\in\Lambda_R}|g\bm{x}\rangle\otimes
   \bigl[(\Gamma^{h(g)}R_g\otimes U_g)|\xi_{\bm{x}}\rangle\bigr].
 \label{eq-intro-enlarged-symmetry}
\end{equation}
Here $R_g$ is the symmetry operator induced by $\rho_V(g)$ on the auxiliary space. The operators $\rho^+(g)$ retain the input data $s,h,\omega$. \Cref{app:morita} gives the explicit auxiliary matrices and their relations.

The resulting zero-dimensional system has the physical symmetries $\rho^+(g)$ together with $n$ additional onsite Clifford generators $iq_j\otimes I$, which anticommute with $m$. The zero-dimensional system with these enlarged symmetry constraints have the same classification as the original $n$-dimensional system.

Numerical diagonalization and comparison with a fixed trivial reference select the subspace used for classification. We count its irreducible blocks under $\rho^+(g)$, the additional Clifford generators, and the reference. Their multiplicities determine the $\mathbb{Z}$ or $\mathbb{Z}_2$ indices through the rules in \cref{app:zero-dimensional-classification}.

This construction gives a classifying-space interpretation of the spectral localizer (\cref{subsec:spectral-localizer}). Spectral-localizer methods combine the Hamiltonian and position operators into an auxiliary zero-dimensional Hamiltonian to compute real-space invariants~\cite{loring2015pseudospectra,loring2017finite}. Crystalline extensions include reflection-based local markers~\cite{cerjan2024crystalline}. Chaou et al.\ develop a systematic treatment of the localizer's zero-dimensional classification and construct its symmetry operators from the spatial symmetries of the original Hamiltonian~\cite{chaou2026simple}.

These symmetry data $(G,\omega,s,h,V)$ encompass spinful and spinless systems with magnetic point-group symmetry~\cite{shiozakiClassificationSurfaceStates2022}, as well as finite spin magnetic point groups, which allow independent spin and spatial transformations and may include nontrivial mixing of point group operations and antiunitary operations~\cite{liuSpinGroupSymmetryMagnetic2022,schiffCrystallographicSpinPoint2025}. They also accommodate superconducting order parameters that transform nontrivially under the symmetry group~\cite{shiozakiClassificationSurfaceStates2022}. \Cref{ex-section} gives examples, and a superconducting example in \cref{ex-spin-pairing} combines these features.

Our GAP package FreeFermionicSPT~\cite{ttiiddeeTtiiddeeFreeFermionicSPT2026} constructs the finite-size zero-dimensional Hamiltonian from the input tight-binding Hamiltonian and computes its topological indices using the symmetry-algebra decomposition of Ref.~\cite{yuanAlgebraFreeFermions2026}. For each example in this paper, the calculation takes at most a few minutes on a standard laptop.

\section{Real-space classifying space Conjecture}
\label{sec-spaces}

This section defines the symmetry algebra $\symalg^W$ and the space of local gapped Hamiltonians, states Kitaev's real-space classification conjecture, and describes the three comparison constructions. Readers interested in the computation results can go directly to the examples in \cref{ex-section}.

\subsection{The symmetry algebra}

Let $G$ be a finite group with an orthogonal action on a finite-dimensional Euclidean space $V$, and set $n=\dim V$. The image of this action is the spatial point group; its kernel may contain internal symmetries. As in Ref.~\cite{yuanAlgebraFreeFermions2026}, the symmetry data include two homomorphisms $s,h:G\to\mathbb{Z}_2$ and a normalized $s$-twisted $2$-cocycle $\omega:G\times G\to U(1)$. The value $s(g)$ specifies whether an internal symmetry operator $U_g$ is linear or antilinear, whereas $h(g)$ specifies whether the physical symmetry commutes or anticommutes with the Hamiltonian. The projective multiplication law is
\begin{equation}\label{eq-symmetry}
 \begin{aligned}
 U_gU_k&=\omega(g,k)U_{gk},\\
 \omega(g,k)\omega(gk,\ell)&={}^{g}\omega(k,\ell)\omega(g,k\ell),
 \end{aligned}
\end{equation}
where ${}^{g}z=z$ for $s(g)=0$ and ${}^{g}z=\overline z$ for $s(g)=1$. Thus $\omega$ is a cocycle representing a class in $H^2(G,U(1)_s)$; the subscript records this action on the coefficients. These data define the real $\mathbb{Z}_2$-graded algebra $\symalg=\C[G,\omega,s]$, with $gi=(-1)^{s(g)}ig$ and degree $|g|=h(g)$. The two gradings $s$ and $h$ are independent.

For a real inner-product space $W$, let $\clifford^W$ denote the real $\mathbb Z_2$-graded Clifford algebra generated by odd elements $\widetilde e(\bm{w})$, linear in $\bm{w}$, with $\widetilde e(\bm{w})^2=|\bm{w}|^2I$. Thus $\clifford^W\cong\clifford^{0,\dim W}$ in our sign convention, where $\clifford^{p,q}$ is the Clifford algebra with $p$ negative generators and $q$ positive generators. If $W$ carries an orthogonal $G$-action, the extension $\symalg^W$ is generated by $\symalg$ and $\clifford^W$. The generators of $\clifford^W$ commute with $i$ and satisfy
\begin{equation}\label{eq-clifford}
 \begin{aligned}
 \{\widetilde e(\bm{v}),\widetilde e(\bm{w})\}&=2\langle \bm{v},\bm{w}\rangle I,\\
 U_g\widetilde e(\bm{w})U_g^{-1}&=(-1)^{h(g)}\widetilde e(g\bm{w}).
 \end{aligned}
\end{equation}
A Hamiltonian with symmetry algebra $\symalg^W$ anticommutes with every generator of $\clifford^W$. Define
\begin{equation}\label{eq-twist}
 \rho_{V'}(g)=(-1)^{s(g)}\rho_V(g).
\end{equation}
Thus $V'$ has the same underlying vector space as $V$, with a twisted action. If $V$ describes the real-space action, $V'$ describes the induced momentum-space action: an antiunitary symmetry introduces an additional momentum reversal. The same twist converts the momentum action back to the real-space action. We write $\widetilde e(\bm{v})$ and $\widetilde e'(\bm{v}')$ for the generators of $\clifford^V$ and $\clifford^{V'}$, respectively.

The adjoint is specified by $i^*=-i$, $g^*=(g)^{-1}$, $\widetilde e(\bm{v})^*=\widetilde e(\bm{v})$, and $(ab)^*=b^*a^*$. Here $(g)^{-1}$ is the inverse in the twisted algebra. For finite $G$ and finite-dimensional $W$, $\symalg^W$ is a finite-dimensional real $\mathbb{Z}_2$-graded $C^*$-algebra. We represent the algebra by operators satisfying these adjoint relations and use the same symbols for the generators and their represented operators. The operator $i$ makes the real Hilbert space into a complex Hilbert space. The relation $g^*=(g)^{-1}$ requires the symmetry operators to be unitary or antiunitary according to $s(g)$, and the generators of $\clifford^W$ are represented by complex-linear self-adjoint operators. For antilinear operators, adjoints are taken with respect to the real part of the Hilbert-space inner product.

\subsection{Locality and Hamiltonians}

Fix a nonempty $G$-invariant lattice $\Lambda\subset V$.\footnote{We assume bounded local geometry and relative density:
\[
 \begin{gathered}
 \sup_{\bm{x}\in\Lambda}\#\{\bm{y}\in\Lambda:|\bm{x}-\bm{y}|\le R\}<\infty\quad(R<\infty),\\
 \sup_{\bm{z}\in V}\operatorname{dist}(\bm{z},\Lambda)<\infty.
 \end{gathered}
\]
The first condition bounds the number of sites in a ball of any fixed radius, uniformly in its centre. The second says that every point of $V$ lies within a fixed distance of some lattice site.}
At each lattice site we place a copy of a complex Hilbert space $M$. It carries a representation of $\symalg$, or of $\symalg^W$ when an action of $\clifford^W$ is also specified. We want to compare systems with different numbers of internal states and allow the addition of trivial states. We therefore choose $M$ large enough to contain countably infinitely many copies of every finite-dimensional irreducible $\mathbb{Z}_2$-graded representation of this algebra. More precisely, $M$ is the Hilbert completion of their orthogonal direct sum. Such a representation is called a complete universe. The grading operator $\varepsilon_M$ is $+I$ on the even part and $-I$ on the odd part, and satisfies $\varepsilon_Ma=(-1)^{|a|}a\varepsilon_M$ for a homogeneous algebra element $a$. It is the designated trivial Hamiltonian.

The single-particle Hilbert space is $\mathcal H=\ell^2(\Lambda;\C)\otimes M$. Here $\ell^2(\Lambda;\C)$ consists of square-summable scalar amplitudes on the lattice. Thus a vector in $\mathcal H$ assigns an internal state $\xi_{\bm{x}}\in M$ to each site, with $\sum_{\bm{x}}\|\xi_{\bm{x}}\|^2<\infty$. If $\delta_{\bm{x}}\in \ell^2(\Lambda;\C)$ is the unit vector at site $\bm{x}$, the spatial action sends $\delta_{\bm{x}}$ to $\delta_{g\bm{x}}$. Combining this permutation with the internal action gives
\begin{equation}\label{eq-hilbert-action}
 \rho(g)(\delta_{\bm{x}}\otimes m)=\delta_{g\bm{x}}\otimes U_gm.
\end{equation}
Extend this action linearly when $s(g)=0$ and antilinearly when $s(g)=1$. It has the same projective multiplier $\omega$ as the internal action. Equivalently, $(\rho(g)\xi)_{\bm{x}}=U_g\xi_{g^{-1}\bm{x}}$. The full reference is $\varepsilon_{\mathcal H}=I\otimes\varepsilon_M$.

A single-particle or BdG Hamiltonian $H$ on $\mathcal H$ is complex-linear, bounded, and self-adjoint. We require a finite propagation bound $r$ such that $H_{\bm{x}\bm{y}}=0$ for $|\bm{x}-\bm{y}|>r$, where $H_{\bm{x}\bm{y}}:M\to M$ is the hopping block from $\bm{y}$ to $\bm{x}$. A nonzero bulk gap means
\begin{equation}\label{eq-bulk-gap}
 H^2\ge\delta^2I\quad\text{for some }\delta>0.
\end{equation}
For each $H$, we require one finite-dimensional reference submodule $M_{\mathrm{fin}}\subset M$, common to all sites, such that $H$ preserves $\ell^2(\Lambda;\C)\otimes M_{\mathrm{fin}}$ and equals the reference on $\ell^2(\Lambda;\C)\otimes M_{\mathrm{fin}}^\perp$. Thus only finitely many internal states per site can differ from the reference, using the same finite subspace $M_{\mathrm{fin}}$ at every lattice site.

\subsection{Classifying spaces and three comparison constructions}
\label{sec-maps}

Write $\realcls_V(\symalg)$ for the set of local gapped Hamiltonians just described, with the specified reference and stabilization. No topology on this set is fixed yet. Its elements need not obey the physical symmetry relations. The group acts by
\begin{equation}\label{eq-hamiltonian-action}
 g\cdot H=(-1)^{h(g)}\rho(g)H\rho(g)^{-1}.
\end{equation}
The fixed-point set consists of symmetric Hamiltonians, for which $\rho(g)H\rho(g)^{-1}=(-1)^{h(g)}H$ for every $g\in G$.

The definition also extends to $\symalg^{W}$. Every $H\in\realcls_V(\symalg^{W})$ must anticommute with the generators of $\clifford^W$:
\[
 \{H,I\otimes\widetilde e(\bm{w})\}=0\qquad(\bm{w}\in W).
\]

Let $S^V=V\cup\{\infty\}$ be the one-point compactification of $V$, with its induced $G$-action. The notation $\Omega^V X$ is intended to mean the based mapping space $\operatorname{Map}_*(S^V,X)$ once a topology on $X$ has been chosen. The three comparison constructions have the following directions:
\begin{equation}\label{eq-three-maps}
\begin{tikzcd}[column sep=small,row sep=large]
 \realcls_V(\symalg) \arrow[dr,"\Sph"'] &&
 \realcls_0(\symalg^{V'}) \arrow[ll,"\Lat"'] \\
 & \Omega^V\realcls_V(\symalg^{V'}) \arrow[ur,"\Disc"'] &
\end{tikzcd}
\end{equation}
The "$0$" in $\realcls_0(\symalg^{V'})$ denotes the $0$-dimensional real orthogonal $G$-representation. We describe the three constructions below.

Inspired by Kitaev's real-space proposal~\cite{alexeikitaevLocalDefinitionsGapped}, we expect these constructions to relate the three classifying spaces. The following conjecture extends this viewpoint to the finite-group symmetry data above.

\begin{conj}[Real-space classifying space]\label{conj-real-space}
For the finite-group data above, the three constructions in \cref{eq-three-maps} admit an equivariant homotopy-theoretic formulation in which each is a weak $G$-homotopy equivalence and $\Disc$ is inverse to $\Sph\circ\Lat$. A weak $G$-homotopy equivalence means a weak homotopy equivalence on the fixed points of every subgroup of $G$.
\end{conj}

We do not prove Conjecture~\ref{conj-real-space}. It would identify the connected components of the symmetric real-space Hamiltonian space with the zero-dimensional classification computed below.

\begin{rk}[Choice of topology]\label{rk-topology}
The choice of topology requires care. The usual operator-norm model may retain onsite occupation data beyond the usual classification. A one-dimensional example retains position-dependent occupation data~\cite[Proposition~6.1]{katoHomotopyTypeUnitary2023}. Related real-space classifications use additional equivalences~\cite[Definition~3.1 and Proposition~3.4]{kubotaControlledTopologicalPhases2017} or a larger class of Hamiltonians~\cite[Definitions~1.1--1.2 and Theorem~1.3]{chungTopologicalClassificationInsulators2026}. We leave the precise classification framework open. This does not prevent us from constructing the finite matrices and computing their zero-dimensional classes; their interpretation as a complete real-space classification is conditional on the conjecture.
\end{rk}

\begin{rk}
In principle, $\realcls_V(\symalg)$ also depends on the choice of lattice $\Lambda \subset V$. However, if Conjecture~\ref{conj-real-space} is true, then all admissible lattices give the same weak $G$-homotopy type. Since we assume Conjecture~\ref{conj-real-space} as the foundation of our computation, we will not write the choice explicitly in the notation $\realcls_V(\symalg)$.
\end{rk}

\paragraph{The sphere construction.}
The map $\Sph$ in \cref{eq-three-maps} is the composite
\begin{equation*}
 \realcls_V(\symalg)
 \xrightarrow{\text{Morita}} \realcls_V(\symalg^{V\oplus V'})
 \xrightarrow{\text{Dirac}} \Omega^V\realcls_V(\symalg^{V'}).
\end{equation*}
The first arrow uses $\mathbb{Z}_2$-graded Morita equivalence. With $V'$ defined by \cref{eq-twist}, the algebraic isomorphism in Ref.~\cite{yuanAlgebraFreeFermions2026} gives
\begin{equation}\label{eq-morita-pair}
 \symalg^{V\oplus V'}\cong\clifford^{n,n}\widehat\otimes\symalg.
\end{equation}
Here $\widehat\otimes$ denotes the $\mathbb{Z}_2$-graded tensor product over $\mathbb{R}$. The algebra $\clifford^{n,n}$ is a $\mathbb{Z}_2$-graded matrix algebra, so $\symalg^{V\oplus V'}$ and $\symalg$ are $\mathbb{Z}_2$-graded Morita equivalent. Apply this equivalence at each site to enlarge $H$ to a Hamiltonian $K$ on $\mathcal K$. The enlargement preserves the gap and propagation of $H$. It also supplies representations of $\clifford^V$ and $\clifford^{V'}$, with generators $\widetilde e(\bm{v})$ and $\widetilde e'(\bm{v}')$, respectively. Generators from the two algebras anticommute with each other. Both $K$ and the enlarged reference $\varepsilon_{\mathcal K}$ anticommute with all these generators. \Cref{app:morita} gives the derivation and the enlargement.

The second arrow uses the generators of $\clifford^V$ to construct the standard sphere Dirac Hamiltonian of Ref.~\cite{yuanAlgebraFreeFermions2026}. In unit-sphere coordinates $v_0^2+|\bm{v}|^2=1$, with the south pole as basepoint, set
\begin{equation}\label{eq-sphere-dirac}
 D_K(v_0,\bm{v})=
 \begin{cases}
 v_0K+\widetilde e(\bm{v}),&v_0\ge0,\\
 -v_0\varepsilon_{\mathcal K}+\widetilde e(\bm{v}),&v_0\le0.
 \end{cases}
\end{equation}
The north pole carries $K$, the south pole carries the reference Hamiltonian, and the two formulas agree at the equator. The generators of $\clifford^{V'}$ anticommute with $D_K(v_0,\bm{v})$ at every sphere point, as required for a Hamiltonian with symmetry algebra $\symalg^{V'}$. The anticommutation relations give a gap of at least $\min(\delta,1)$, without flattening $H$.

\paragraph{Discretization.}
Let $F:S^V\to\realcls_V(\symalg^{V'})$ be a based map, with $F(\infty)=\varepsilon_{\mathcal K}$. Choose a slowly varying $G$-equivariant map $\theta:V\to S^V$ that equals $\infty$ outside a bounded region. The parameter assigned to each lattice site is given by the composite
\[
 \Lambda\hookrightarrow V\xrightarrow{\theta}S^V
 \xrightarrow{F}\realcls_V(\symalg^{V'}).
\]
For a hopping block from $\bm{y}$ to $\bm{x}$, the two endpoints may select different Hamiltonians. We combine them by the endpoint average
\begin{equation}\label{eq-endpoint-average}
 [\Disc_\theta(F)]_{\bm{x}\bm{y}}=\tfrac12\bigl(F(\theta(\bm{x}))_{\bm{x}\bm{y}}+F(\theta(\bm{y}))_{\bm{x}\bm{y}}\bigr).
\end{equation}
Although every $F(p)$ is gapped, combining hopping blocks from different parameter values can close the gap. The parameters must therefore vary slowly across the hopping range. A concrete radial parameterization and a finite upper-hemisphere implementation are given in \cref{app:parameterization}.

The endpoint average is self-adjoint and anticommutes with every generator of $\clifford^{V'}$. Equivariance of $F$ and $\theta$ gives the physical symmetry of the Hamiltonian. Outside the region where $\theta\ne\infty$, enlarged by one propagation radius, the operator is exactly the reference. Its non-reference part therefore lies in a finite-dimensional subspace. When it is gapped, it defines a zero-dimensional Hamiltonian with symmetry algebra $\symalg^{V'}$. The symmetry representation on this finite space includes both the internal action and the permutations of lattice sites.

\paragraph{Lattice realization.}
The numerical algorithm uses only $\Disc\circ\Sph$ and does not evaluate $\Lat$. To describe $\Lat$, start from a zero-dimensional mass $m\in\realcls_0(\symalg^{V'})$ and form the standard sphere Dirac Hamiltonian. Write $S^{V'}=\{(v_0,\bm{v}):v_0^2+|\bm{v}|^2=1\}$ and set
\begin{equation}\label{eq-dirac-momentum-sphere}
 D_m(v_0,\bm{v})=
 \begin{cases}
 v_0m+\widetilde e'(\bm{v}),&v_0\ge0,\\
 -v_0\varepsilon_M+\widetilde e'(\bm{v}),&v_0\le0.
 \end{cases}
\end{equation}
The north pole carries $m$, and the south pole carries the reference Hamiltonian. Here the sphere describes compactified momentum space, with the action on $V'$. To obtain $\Lat(m)$, $D_m$ must be converted into a local gapped real-space Hamiltonian on the given lattice $\Lambda$. The rough idea is to apply an inverse Fourier transform.

\subsection{Morita equivalence for the sphere construction}
\label{app:morita}

We construct the $\mathbb Z_2$-graded algebra isomorphism
\begin{equation}
 \Phi:\symalg^{V\oplus V'}\xrightarrow{\cong}
 \operatorname{End}_{\R}(S)\widehat\otimes_{\R}\symalg
 \cong\clifford^{n,n}\widehat\otimes_{\R}\symalg.
 \label{eq-morita-separated-algebra}
\end{equation}
Here $n=\dim V$ and $\rho_{V'}(g)=(-1)^{s(g)}\rho_V(g)$. Ref.~\cite{yuanAlgebraFreeFermions2026} obtained this isomorphism by the general separation-of-variables theorem and cancellation of the cocycle corrections. Here we provide a direct construction.

Let $S=\Lambda^*V$ be the real exterior algebra, with its induced inner product and $\mathbb Z_2$-grading $\Gamma|_{\Lambda^kV}=(-1)^kI$. For $\bm{v}\in V$, define exterior multiplication by $a(\bm{v})\alpha=\bm{v}\wedge\alpha$ for $\alpha\in S$. Its adjoint $\iota(\bm{v})=a(\bm{v})^*$ is contraction:
\begin{equation}
 \begin{multlined}
 \iota(\bm{v})(\bm{w}_1\wedge\cdots\wedge \bm{w}_k)\\
 =\sum_{j=1}^k(-1)^{j-1}\langle \bm{v},\bm{w}_j\rangle
 \bm{w}_1\wedge\cdots\wedge\widehat{\bm{w}_j}\wedge\cdots\wedge \bm{w}_k.
 \end{multlined}
 \label{eq-morita-contraction}
\end{equation}
Here the hat denotes omission, and $\iota(\bm{v})\,1=0$. Set
\begin{equation}
 c(\bm{v})=a(\bm{v})+\iota(\bm{v}),\qquad q(\bm{v})=\iota(\bm{v})-a(\bm{v}).
 \label{eq-morita-exterior-generators}
\end{equation}
They are odd operators with $c(\bm{v})^2=|\bm{v}|^2I$, $q(\bm{v})^2=-|\bm{v}|^2I$, and $\{c(\bm{v}),q(\bm{w})\}=0$. They generate $\operatorname{End}_{\R}(S)$, realizing $\clifford^{n,n}$ on its $2^n$-dimensional real $\Z_2$-graded simple module $S$; see Ref.~\cite[Sec.~1.4]{AST_2009__327__131_0}.

The spatial action induces $R_g=\Lambda^*\rho_V(g)$ on $S$, explicitly
\begin{equation}
 \begin{multlined}
 R_g(\bm{w}_1\wedge\cdots\wedge \bm{w}_k)\\
 =\rho_V(g)\bm{w}_1\wedge\cdots\wedge\rho_V(g)\bm{w}_k,\qquad R_g(1)=1.
 \end{multlined}
 \label{eq-morita-exterior-action}
\end{equation}
Identifying the underlying vector spaces of $V$ and $V'$, define $\Phi$ on generators by
\begin{equation}
 \begin{aligned}
 \Phi(i)&=I\widehat\otimes i,&
 \Phi(g)&=R_g\widehat\otimes g,\\
 \Phi(\widetilde e(\bm{v}))&=c(\bm{v})\widehat\otimes1,&
 \Phi(\widetilde e'(\bm{v}))&=q(\bm{v})\widehat\otimes i.
 \end{aligned}
 \label{eq-morita-generator-map}
\end{equation}
The $\mathbb Z_2$-graded tensor product supplies the sign $(-1)^{h(g)}$, while $gi=(-1)^{s(g)}ig$ gives the twist in $V'$. Together with $R_gR_k=R_{gk}$, these identities verify the defining relations. The images generate the target: the Clifford images and $\Phi(i)$ give the full matrix factor, and $(R_g^{-1}\widehat\otimes1)\Phi(g)=I\widehat\otimes g$. Thus $\Phi$ is surjective. Ordered Clifford monomials give the source dimension at most $2^{2n}\dim_{\R}\symalg$, equal to the target dimension, so $\Phi$ is an isomorphism.

To enlarge the system, take the $\mathbb Z_2$-graded tensor product $\mathcal K=S\widehat\otimes_{\R}\mathcal H$. Its natural $\operatorname{End}_{\R}(S)\widehat\otimes_{\R}\symalg$ action gives an action of $\symalg^{V\oplus V'}$ through $\Phi$. For $X\in\operatorname{End}_{\R}(S)$, $\xi\in\mathcal H$, and homogeneous $a\in\symalg$ and $u\in S$, the tensor-product action is
\begin{equation}
 (X\widehat\otimes a)(u\widehat\otimes\xi)
 =(-1)^{|a||u|}Xu\widehat\otimes a\xi.
 \label{eq-morita-super-tensor-action}
\end{equation}
Equivalently, on the underlying ordinary tensor product $S\otimes_{\R}\mathcal H$, $X\widehat\otimes a$ is represented by $X\Gamma^{|a|}\otimes a$. In particular, the enlarged physical symmetry is $\rho^+(g)=\Gamma^{h(g)}R_g\otimes\rho(g)$. The grading on $S$ is $\Gamma$, so the Hamiltonian and reference are $K=\Gamma\otimes H$ and $\varepsilon_{\mathcal K}=\Gamma\otimes\varepsilon_{\mathcal H}$.

\section{Computational procedure}
\label{sec:algorithm}

Under Conjecture~\ref{conj-real-space}, the class of a symmetric Hamiltonian $H$ is given by the zero-dimensional class of $(\Disc\circ\Sph)(H)$. First choose a slowly varying $G$-equivariant map with a parameter
\[
 \theta_R:V\longrightarrow S^V,\qquad R>0,
\]
where $R$ controls the spatial size of $(\Disc_{\theta_R}\circ\Sph)(H)$. Each $\theta_R$ equals $\infty$ outside a finite region, and increasing $R$ slows the parameter variation.

For the numerical calculations we use the upper-hemisphere parameterization $\theta_R^+$ given in \cref{app:parameterization}, and \cref{subsec:spectral-localizer} explains why the upper hemisphere alone suffices. Let $m$ denote the resulting finite-size zero-dimensional Hamiltonian. When it is gapped, we compute its class by flattening its spectrum and applying the method in \cref{app:zero-dimensional-classification}.

For a semisimple $\mathbb Z_2$-graded algebra $A$, its Grothendieck group $\mathcal N(A)$ is freely generated by the isomorphism classes of irreducible $\mathbb Z_2$-graded $A$-modules. The zero-dimensional classification group is
\[
 \coker\!\left[
 \mathcal N(\clifford^{0,1}\widehat\otimes\symalg^{V'})
 \longrightarrow\mathcal N(\symalg^{V'})
 \right],
\]
where the arrow forgets the generator of the added $\clifford^{0,1}$ factor. The flattened Hamiltonian and its reference determine a module, as described in \cref{app:zero-dimensional-classification}.

Our GAP package FreeFermionicSPT~\cite{ttiiddeeTtiiddeeFreeFermionicSPT2026} implements this procedure. It provides symmetry templates for ordinary and magnetic point groups, finite spin point groups, and superconducting pairing, together with symmetry-algebra decompositions and Clifford restriction maps. Given a real-space Hamiltonian, its symmetry action and a reference, the package constructs the finite-size zero-dimensional Hamiltonian and returns its classification group and indices relative to the reference. Source code, tutorials, and worked examples are available in the accompanying GitHub repository.

\subsection{The finite-size zero-dimensional Hamiltonian}
\label{app:parameterization}

\subsubsection{Choice of parameterization}

We choose $\theta_R^+:V\longrightarrow S^V$ to take values in the upper hemisphere. The parameterization is not unique; different choices may only slightly affect numerical efficiency. Write $S^V$ as the unit sphere $v_0^2+|\bm{v}|^2=1$. Fix $R>0$. For $|\bm{x}|<R$, set
\begin{equation}
 \begin{aligned}
 v_0(\bm{x})&=\cos\!\left(\frac{\pi|\bm{x}|}{2R}\right),\\
 \bm{v}(\bm{x})&=\sin\!\left(\frac{\pi|\bm{x}|}{2R}\right)\frac{\bm{x}}{|\bm{x}|}\quad(\bm{x}\neq\bm{0}),\\
 \bm{v}(\bm{0})&=\bm{0}.
 \end{aligned}
 \label{eq-radial-sphere-coordinates}
\end{equation}
On the boundary and outside the ball, set
\begin{equation}
 v_0(\bm{x})=0,\qquad \bm{v}(\bm{x})=\frac{\bm{x}}{|\bm{x}|},
 \qquad |\bm{x}|\ge R.
 \label{eq-radial-exterior}
\end{equation}
These formulas directly define
\begin{equation}
 \theta_R^+:V\longrightarrow S^V,\qquad
 \theta_R^+(\bm{x})=(v_0(\bm{x}),\bm{v}(\bm{x})).
 \label{eq-radial-parameterization}
\end{equation}
The origin maps to the north pole. Along each ray, the sphere point reaches the equator at radius $R$ and stays there. Its equatorial value retains the direction $\bm{x}/|\bm{x}|$. The map is continuous and $G$-equivariant, and its image lies in the closed upper hemisphere.

\subsubsection{The finite-size zero-dimensional system}

Endpoint averaging of the upper-hemisphere Hamiltonians, as in \cref{eq-endpoint-average}, gives the matrix elements
\begin{equation}
 \frac{v_0(\bm{x})+v_0(\bm{y})}2K_{\bm{x}\bm{y}}
       +\delta_{\bm{x}\bm{y}}\widetilde e(\bm{v}(\bm{x})),
 \qquad \bm{x},\bm{y}\in\Lambda.
 \label{eq-radial-endpoint-hamiltonian}
\end{equation}
Although $v_0(\bm{x})=0$ for $|\bm{x}|\ge R$, such a site can still couple to a site $\bm{y}$ with $|\bm{y}|<R$ through $v_0(\bm{y})K_{\bm{x}\bm{y}}/2$. To eliminate this hopping across radius $R$, we replace the arithmetic mean by the geometric mean and define
\begin{equation}
 \begin{gathered}
 m_{\bm{x}\bm{y}}=\sqrt{v_0(\bm{x})v_0(\bm{y})}\,K_{\bm{x}\bm{y}}
          +\delta_{\bm{x}\bm{y}}\widetilde e(\bm{v}(\bm{x})),\\
 \bm{x},\bm{y}\in\Lambda,\quad |\bm{x}|,|\bm{y}|<R.
 \end{gathered}
 \label{eq-radial-finite-hamiltonian}
\end{equation}
For sufficiently slow parameter variation (large enough $R$), the two constructions give the same zero-dimensional class. A symmetry-preserving gapped deformation connects them. %

The replacement is optional: direct endpoint averaging only requires the larger region $|\bm{x}|\le R+r$, where $r$ bounds the propagation of $K$.

With the weights in \cref{eq-radial-finite-hamiltonian}, all sites with $|\bm{x}|\ge R$ have only the decoupled onsite blocks $\widetilde e(\bm{x}/|\bm{x}|)$. The finite matrix $m$ therefore acts on the sites $\{\bm{x}\in\Lambda:|\bm{x}|<R\}$. This set is $G$-invariant. The physical symmetry acts by both the internal operators and the permutations of these sites, and the generators of $\clifford^{V'}$ anticommute with $m$. The reference is the restriction of $\varepsilon_{\mathcal K}$ to this finite space. Whenever $m$ is gapped, it defines a zero-dimensional Hamiltonian with symmetry algebra $\symalg^{V'}$.

We compute the class of $m$ using \cref{app:zero-dimensional-classification}. In practice, we can repeat the calculation for increasing $R$ and examine the numerical stability of the resulting class.

\subsection{Reading the class of a zero-dimensional Hamiltonian}
\label{app:zero-dimensional-classification}

The starting point for the zero-dimensional calculation is the correspondence between a $K$-group and Hamiltonians obtained by changing the sign of a reference mass term. We first recall what this correspondence means and how it depends on the reference. We then explain how to compute its inverse for a given Hamiltonian: flatten the Hamiltonian and deform it to a representative that differs from the reference only by a sign on some $\mathbb{Z}_2$-graded submodules. Throughout this subsection the algebra is the full $\symalg^{V'}$ of \cref{sec-spaces}. Its action includes the physical symmetries and the generators of $\clifford^{V'}$.

\subsubsection{The \texorpdfstring{$K$}{K}-group and the reference}
\label{zd-algebraic-classification}

Recall that $\mathcal N(\symalg^{V'})$ is the free abelian group generated by the isomorphism classes of irreducible $\mathbb{Z}_2$-graded $\symalg^{V'}$-modules. Direct sum becomes addition, and formal differences are allowed. The $K$-group used in \cref{sec:algorithm} is the cokernel of the restriction map:
\begin{equation}
 \begin{gathered}
 \mathcal N(\clifford^{0,1}\widehat\otimes\symalg^{V'})
 \xrightarrow{\ \mathrm{res}\ }
 \mathcal N(\symalg^{V'}),\\
 \coker(\mathrm{res})=
 \frac{\mathcal N(\symalg^{V'})}{\operatorname{im}(\mathrm{res})}.
 \end{gathered}
 \label{eq-zd-classification-quotient}
\end{equation}
The map $\mathrm{res}$ forgets the action of the added $\clifford^{0,1}$ factor and views the same $\mathbb Z_2$-graded vector space as an $\symalg^{V'}$-module. The quotient therefore identifies certain combinations of $\mathbb Z_2$-graded simple modules with zero. To interpret this group as a classification of Hamiltonians, we must specify a reference Hamiltonian.

Let $W$ be a unitary reference module, with grading operator $\varepsilon_W$. This operator is $+I$ on the even part and $-I$ on the odd part, and satisfies $\varepsilon_W^*=\varepsilon_W$ and $\varepsilon_W^2=I$. It is the Hamiltonian designated as trivial. We allow reference stabilization as in \cref{sec-spaces}: $W$ may be enlarged by reference modules from a complete universe containing arbitrarily many copies of every finite $\mathbb{Z}_2$-graded simple type.

Choose a $\mathbb{Z}_2$-graded submodule $F\subset W$. Change the sign of the reference Hamiltonian on $F$, while keeping the reference on its orthogonal complement:
\begin{equation}
 m_F=(-\varepsilon_W|_F)\oplus(\varepsilon_W|_{F^\perp})
     =(I-2P_F)\varepsilon_W,
 \label{eq-zd-flip-representative}
\end{equation}
where $P_F$ is the orthogonal projection onto $F$. Because $F$ is a $\mathbb{Z}_2$-graded submodule, $P_F$ commutes with the full algebra action and with $\varepsilon_W$. Thus $m_F$ satisfies all Hamiltonian symmetry relations and has square $I$.

This construction is the \emph{changing the sign of the mass term} correspondence of Ref.~\cite{yuanAlgebraFreeFermions2026}:
\begin{equation}
 [(F,\varepsilon_W|_F)]
     \longmapsto [m_F]_{\varepsilon_W}.
 \label{eq-zd-sign-change-map}
\end{equation}
The brackets on the left denote a class in the algebraic quotient in \cref{eq-zd-classification-quotient}; those on the right denote a gapped homotopy class with the chosen reference stabilization. If we take $W$ to be the complete reference universe, this map is an isomorphism between the $K$-group and the stable zero-dimensional classification. The restriction relations have a physical meaning: if a reference module extends to $\clifford^{0,1}\widehat\otimes\symalg^{V'}$, its sign reversal can be undone by a gapped deformation induced by the extra mass term.

We write $\ind(m;\varepsilon_W)$ for the algebraic label of a Hamiltonian $m$ relative to the reference. In particular, $\ind(\varepsilon_W;\varepsilon_W)=0$, and \cref{eq-zd-sign-change-map} gives $\ind(m_F;\varepsilon_W)=[(F,\varepsilon_W|_F)]$. The symmetry data determine the abstract classification group, but assigning an element of that group to a particular Hamiltonian depends on the reference: changing the designated trivial Hamiltonian can change the label of the same physical system. Thus the label is a property of the pair $(m,\varepsilon_W)$, and the grading used to count flipped modules is the \emph{reference} grading.

For a Hamiltonian already of the form $m_F$, we only need to decompose $F$, with its reference $\mathbb{Z}_2$-grading, into simple modules and read its class in the quotient. Flips add under direct sum; the quotient relations turn their multiplicities into $\mathbb{Z}$ or $\mathbb{Z}_2$. For a general input, the remaining problem is to find such a flip representative.

\subsubsection{The computational steps}
\label{zd-reference-flattening}

Let $m$ denote the given zero-dimensional Hamiltonian on $W$. In the computational procedure, it is the finite-size Hamiltonian constructed in \cref{app:parameterization}. We assume $m=m^*$ and $m^2\geq\delta_m^2I$ for some $\delta_m>0$. Write $\rho(a)$ for the action of a homogeneous element $a\in\symalg^{V'}$, and $|a|\in\{0,1\}$ for its degree. The Hamiltonian and the reference satisfy
\begin{equation}
 \varepsilon_W\rho(a)=(-1)^{|a|}\rho(a)\varepsilon_W,
 \qquad
 m\rho(a)=(-1)^{|a|}\rho(a)m.
 \label{eq-zd-mass-relations}
\end{equation}
Both $m$ and $\varepsilon_W$ are complex-linear. The algebra action is understood over $\mathbb R$, so these relations also include the antilinear symmetries.

To invert the sign-change correspondence for this input, we perform two steps:
\begin{enumerate}
 \item Flatten $m$ to a Hamiltonian $\widehat m$ with eigenvalues $+1$ and $-1$, without changing its gapped homotopy class.
 \item Find a $\mathbb{Z}_2$-graded reference submodule $F$ and a gapped deformation from $\widehat m$ to $m_F$, which equals $-\varepsilon_W$ on $F$ and $\varepsilon_W$ on $F^\perp$.
\end{enumerate}
Once these steps are completed, the answer is $[(F,\varepsilon_W|_F)]$ in the algebraic quotient. The class depends only on that quotient element; a particular choice of the flipped submodule need not be unique.

The first step is standard Hermitian linear algebra. Spectral flattening replaces each positive eigenvalue of $m$ by $+1$ and each negative eigenvalue by $-1$:
\begin{equation}
 \widehat m=\sgn(m),
 \qquad \widehat m^*=\widehat m,\qquad \widehat m^2=I.
 \label{eq-zd-flattening}
\end{equation}
For example, a unitary diagonalization $m=U\operatorname{diag}(x_j)U^*$ gives $\widehat m=U\operatorname{diag}(\sgn(x_j))U^*$. The real odd function $\sgn$ preserves \cref{eq-zd-mass-relations}, including the antilinear symmetries. Moreover, $(1-t)m+t\widehat m$ stays gapped for $0\leq t\leq1$: an eigenvalue $x\neq0$ becomes $(1-t)x+t\sgn(x)$ and retains its sign. The gap along this path is at least $\min(\delta_m,1)$.

The second step needs an additional construction. Although $\widehat m$ is a grading of the underlying $\symalg^{V'}$-module, it can mix the submodules used to describe the reference and need not commute with $\varepsilon_W$. It is therefore not yet a representative whose only change is a reference sign reversal. The comparison operator below constructs a suitable representative.

\subsubsection{Finding a flip representative}
\label{zd-comparison}

A similar method is used in Ref.~\cite[Sec.~4]{bourneKOvaluedSpectralFlow2022}.
To find the part that can return to the reference, consider the direct interpolation
\begin{equation}
 A_t=(1-t)\widehat m+t\varepsilon_W.
 \label{eq-zd-reference-path}
\end{equation}
Its square is
\begin{equation}
 A_t^2=\bigl((1-t)^2+t^2\bigr)I+2t(1-t)C,
 \qquad
 C=\frac{\varepsilon_W\widehat m+\widehat m\varepsilon_W}{2}.
 \label{eq-zd-comparison}
\end{equation}
The first term is positive, therefore, $C$ measures whether this interpolation closes the gap.

The operator $C$ is self-adjoint and commutes with $\varepsilon_W$, $\widehat m$, and the full action of $\symalg^{V'}$. The last assertion follows because each product $\varepsilon_W\widehat m$ and $\widehat m\varepsilon_W$ acquires two copies of the sign in \cref{eq-zd-mass-relations}. Furthermore,
\begin{equation}
 (\varepsilon_W+\widehat m)^2=2(I+C),
 \qquad
 (\varepsilon_W-\widehat m)^2=2(I-C).
 \label{eq-zd-comparison-endpoints}
\end{equation}
Both squares are nonnegative, so $\spec(C)\subset[-1,1]$.

For an eigenvalue $c$ of $C$, let $E_c=\ker(C-cI)$ be its eigenspace. This is a $\mathbb{Z}_2$-graded $\symalg^{V'}$-submodule with the reference grading. On $E_c$, \cref{eq-zd-comparison} gives
\begin{equation}
 A_t^2=\bigl(1-2t(1-t)(1-c)\bigr)I,
 \qquad
 \min_{0\leq t\leq1} A_t^2=\frac{1+c}{2}I.
 \label{eq-zd-eigenspace-gap}
\end{equation}
Consequently every part with $c>-1$ can return to the reference along this interpolation. At $c=\pm1$, \cref{eq-zd-comparison-endpoints} gives
\begin{equation}
 \begin{gathered}
 \widehat m=\varepsilon_W\ \hbox{on }E_1,
 \qquad
 \widehat m=-\varepsilon_W\ \hbox{on }E_{-1},\\
 E_{-1}=\ker(\widehat m+\varepsilon_W).
 \end{gathered}
 \label{eq-zd-exact-flip-space}
\end{equation}
For a fixed finite matrix, leave $E_{-1}$ unchanged and apply the interpolation on its complement. The resulting path is gapped and ends at $m_F$ with $F=E_{-1}$. This completes the second computational step in exact arithmetic: the input class is the class of the $\mathbb{Z}_2$-graded reference module $E_{-1}$ in \cref{eq-zd-classification-quotient}.

\subsubsection{Some numerical subtleties}
\label{zd-spectral-window}

In exact arithmetic one can use $E_{-1}$. Numerical arithmetic does not reliably distinguish an eigenvalue exactly equal to $-1$ from one extremely close to it. Instead choose a spectral cut
\begin{equation}
 \begin{gathered}
 -1<\lambda<1,\qquad \lambda\notin\spec(C),\\
 P_\lambda=\mathbf 1_{[-1,\lambda]}(C),
 \qquad
 W_\lambda=\operatorname{im}P_\lambda,
 \end{gathered}
 \label{eq-zd-window}
\end{equation}
where $\spec(C)$ is the set of spectra of $C$. The operator $\mathbf 1_{[-1,\lambda]}(C)$ is the orthogonal projection onto all eigenspaces of $C$ with eigenvalue at most $\lambda$. Its image is $W_\lambda$.

Because $C$ commutes with the full symmetry algebra and the reference, $W_\lambda$ is again a $\mathbb{Z}_2$-graded reference module. More explicitly,
\begin{equation}
 W_\lambda=E_{-1}\oplus
 \bigoplus_{-1<c\leq\lambda}E_c.
 \label{eq-zd-window-sum}
\end{equation}
We show directly that flipping the reference on this window gives a Hamiltonian equivalent to the input. Define
\begin{equation}
 \widehat m_\lambda=(I-2P_\lambda)\varepsilon_W,
 \qquad
 A_{\lambda,t}=(1-t)\widehat m+t\widehat m_\lambda.
 \label{eq-zd-window-path}
\end{equation}
The endpoint $\widehat m_\lambda$ equals $-\varepsilon_W$ on $W_\lambda$ and $\varepsilon_W$ on its complement. Since $P_\lambda$ commutes with $\widehat m$ and $\varepsilon_W$,
\[
 A_{\lambda,t}^2=\bigl((1-t)^2+t^2\bigr)I
       +2t(1-t)(I-2P_\lambda)C.
\]
On $W_\lambda$, the last operator $(I-2P_\lambda)C=-C$ is bounded below by $-\lambda I$; on the complement it equals $C$ and is bounded below by $\lambda I$. Therefore
\begin{equation}
 A_{\lambda,t}^2\geq
 \bigl[1-2t(1-t)(1+|\lambda|)\bigr]I
 \geq\frac{1-|\lambda|}{2}I>0.
 \label{eq-zd-window-path-gap}
\end{equation}
All mass relations are preserved, and the path connects $\widehat m$ to $\widehat m_\lambda$ without closing the gap. If a path of flattened Hamiltonians is desired, apply spectral flattening to $A_{\lambda,t}$ along the path. The sign-change correspondence in \cref{eq-zd-sign-change-map} therefore gives
\begin{equation}
 \begin{aligned}
 \ind(m;\varepsilon_W)
   &=[(W_\lambda,\varepsilon_W|_{W_\lambda})]\\
   &=[(E_{-1},\varepsilon_W|_{E_{-1}})]
   \quad\hbox{in }\coker(\mathrm{res}).
 \end{aligned}
 \label{eq-zd-index}
\end{equation}

\subsubsection{Read the topological indices}
\label{zd-module-coordinates}

Let $S_1,\ldots,S_p$ be the irreducible $\mathbb{Z}_2$-graded $\symalg^{V'}$-modules obtained from our GAP package~\cite{ttiiddeeTtiiddeeFreeFermionicSPT2026}, together with their dimensions and the corresponding simple-algebra types. The sign-change correspondence respects this block decomposition, and the classification for each simple type has already been computed in Ref.~\cite{yuanAlgebraFreeFermions2026}.

Decompose the window with its reference $\mathbb{Z}_2$-grading:
\begin{equation}
 (W_\lambda,\varepsilon_W|_{W_\lambda})
   \cong\bigoplus_{i=1}^p S_i^{\oplus\nu_i}.
 \label{eq-zd-window-multiplicities}
\end{equation}
If $W_{\lambda,i}$ denotes the sum of all copies of $S_i$ in this window, its multiplicity is
\begin{equation}
 \nu_i=\frac{\dim_{\mathbb R}W_{\lambda,i}}
                 {\dim_{\mathbb R}S_i}.
 \label{eq-zd-dimension-normalization}
\end{equation}
The denominator is the dimension of the simple module in the represented space, so the repeated internal degrees of freedom are counted with the correct normalization.

For each graded-simple algebra factor, read the index from its known type:
\begin{itemize}
 \item $\mathbb Z$: the two oppositely graded simple types represent $+1$ and $-1$ after fixing the generator convention. If their multiplicities are $\nu_+$ and $\nu_-$, the index is $\nu_+-\nu_-$.
 \item $\mathbb Z_2$: the graded simple module represents the generator, so its multiplicity $\nu$ gives the index $\nu\bmod2$.
 \item $0$: this factor contributes zero.
\end{itemize}
All multiplicities use the reference grading. Different algebra factors give separate coordinates, even when they have the same simple-algebra type.

\subsubsection{Numerical evaluation and cost}
\label{zd-numerical-cost}

For a given finite input, the general procedure is therefore to flatten $m$, form $C$, choose a separated spectral cut, identify the $\mathbb{Z}_2$-graded reference module $W_\lambda$, and read its multiplicities block by block as in \cref{zd-module-coordinates}. The symmetry decomposition, simple-module types and dimensions, and generator conventions can be reused for all Hamiltonians on the same represented reference space.

For dense matrices of order $N=\dim_{\mathbb C}W$, the two main spectral computations are flattening $m$ and diagonalizing $C$ to obtain the window $W_\lambda$. Each costs $O(N^3)$ using a standard Hermitian eigensolver. In a basis where $\varepsilon_W$ is diagonal, forming $C$ costs $O(N^2)$, and reading integers or parities from the known module multiplicities is inexpensive. Thus the dense computation has cubic complexity, with both spectral steps contributing to the cost.

Symmetry can split these computations into independent smaller blocks and remove repeated simple-module factors. If the reduced matrix sizes are $N_i$, the dense spectral cost becomes $\sum_i O(N_i^3)$. The cost of constructing and applying the decomposition must also be included; when the representation is fixed, the decomposition can be reused across different Hamiltonians. Exploiting the structure and sparsity of the reduced coefficient matrices can further reduce the computational cost.

\section{Relation to the spectral localizer}
\label{subsec:spectral-localizer}

The spectral localizer is obtained by discretizing the upper hemisphere of the sphere construction. For $v_0>0$, removing the positive scalar factor in \cref{eq-sphere-dirac} gives
\begin{equation}\label{eq-sphere-localizer}
 \frac{D_K(v_0,\bm{v})}{v_0}=K+\widetilde e\!\left(\frac{\bm{v}}{v_0}\right).
\end{equation}
Assigning $\bm{v}/v_0=\kappa \bm{x}$ to site $\bm{x}$ gives the standard spectral localizer~\cite{loring2015pseudospectra,loring2017finite,chaou2026simple}:
\begin{equation}\label{eq-linear-localizer}
 (L_\kappa)_{\bm{x}\bm{y}}=K_{\bm{x}\bm{y}}+\kappa\delta_{\bm{x}\bm{y}}\widetilde e(\bm{x}),
 \qquad \kappa>0.
\end{equation}

We show the equivalence of two zero-dimensional constructions. In $L_\kappa$, the position term $\kappa\widetilde e(\bm{x})$ dominates the Hamiltonian $K$ as $|\bm{x}|\to\infty$. In $\Disc_{\theta_R}(\Sph(H))$, the radial parameter reaches the equator with onsite term $\widetilde e(\bm{x}/|\bm{x}|)$. On the lower hemisphere of \cref{eq-sphere-dirac}, this term deforms to the reference $\varepsilon_{\mathcal K}$. A slowly varying $\theta_R$ makes this a gradual spatial interpolation by onsite Hamiltonians.
The outer onsite blocks corresponding to the lower hemisphere can then be rotated back to $\widetilde e(\bm{x}/|\bm{x}|)$. This rotation stays gapped because the reference anticommutes with the position Clifford term. Together with positive rescaling of the upper-hemisphere weights, this gives the same zero-dimensional class as the spectral localizer. The deformation preserves the physical symmetries and anticommutation with the generators of $\clifford^{V'}$.

For the numerical implementation, we use an upper-hemisphere map $\theta_R^+:V\to S^V$ directly, and apply it to $D_K$. The boundary and exterior retain the radial Clifford blocks $\widetilde e(\bm{x}/|\bm{x}|)$. With the hopping weights specified in \cref{app:parameterization}, these blocks decouple from the finite core, so only the core Hamiltonian $m$ needs to be assembled in the calculation.

\section{Examples}\label{ex-section}

For each example we specify the Hamiltonian and its reference, the physical symmetries and symmetry algebra, and the resulting classification. We use $\sigma_x,\sigma_y,\sigma_z$ for the Pauli matrices and $\mathsf K$ for complex conjugation. The finite-size Hamiltonian at radius $R$ is denoted by $m_R$, and $\ind(m_R)=\ind(m_R;\varepsilon_{\mathcal K})$ as an element in the classification $K$-group, as in \cref{app:zero-dimensional-classification}. We use the parameterization in \cref{app:parameterization}.

Sites have integer labels $\bm{n}\in\mathbb Z^d$ and physical positions $\bm{x}_{\bm{n}}=\sum_jn_j\bm{a}_j$. Thus $\bm{n}\mapsto \bm{n}+\bm{e}_j$ is a physical displacement by $\bm{a}_j$. The hopping blocks are $H_{\bm{n},\bm{n}'}$.

For the models below, we scan the integer radii $R=1,\ldots,8$. We denote by $R_{\mathrm{stab}}$ the smallest sampled radius such that $\ind(m_R)=\ind(m_8)$ for $1\le R\le 8$. This demonstrates numerical stability over the sampled integer radii.

\subsection{SSH and Majorana chains}\label{ex-chains}

\paragraph{Hamiltonian.}
On $\Lambda=\mathbb Z$, with two internal states per site, consider
\begin{equation}
 \begin{aligned}
 H_{n,n}&=t_1\sigma_z,\\
 H_{n,n+1}&=\frac{t_2}{2}(\sigma_z+i\sigma_y),\\
 H_{n+1,n}&=H_{n,n+1}^*,
 \end{aligned}
 \label{ex-chain-hopping}
\end{equation}
and reference $\varepsilon=\sigma_z$. The onsite rotation $e^{-i\pi\sigma_y/4}$ converts this into the usual intracell and intercell hopping form of the SSH chain~\cite{su1979solitons}. We take $(t_1,t_2)=(0,1)$; the reference corresponds to $(1,0)$. Both endpoints have bulk gap $1$.

\paragraph{Symmetries and symmetry algebra.}
The SSH chain is classified with the onsite unitary chiral symmetry $S$, represented by $U_S=\sigma_x$, satisfying $U_S^2=I$ and $U_S H U_S^{-1}=-H$. Thus $s(S)=0$ and $h(S)=1$, giving class AIII. Alternatively, interpreting the two components as Nambu degrees of freedom and imposing the onsite particle-hole symmetry $C$, represented by $U_C=\sigma_x\mathsf K$, gives the Majorana chain in class D~\cite{kitaevUnpairedMajoranaFermions2001}. Here $U_C^2=I$, $U_C H U_C^{-1}=-H$, and $s(C)=h(C)=1$. These are two different choices of symmetry constraints on the same Hamiltonian. Their symmetry algebras are
\begin{equation}
 \symalg(\mathrm{AIII})\cong\mathbb C\widehat\otimes\clifford^{0,1},
 \qquad
 \symalg(\mathrm D)\cong\clifford^{0,2}.
 \label{ex-chain-source-algebras}
\end{equation}
Both symmetries fix spatial sites. Consequently, the action on the one-dimensional $V'$ is trivial for $S$ and multiplication by $-1$ for $C$. The zero-dimensional target algebras are
\begin{equation}
 \symalg^{V'}(\mathrm{AIII})\cong\mathbb C\widehat\otimes\clifford^{0,2},
 \qquad
 \symalg^{V'}(\mathrm D)\cong M_2(\clifford^{0,1}).
 \label{ex-chain-target-algebras}
\end{equation}

\paragraph{Classification.}
The corresponding classification groups are $\mathbb Z$ and $\mathbb Z_2$. For every tested integer radius $R=2,\ldots,8$, the computation gives
\begin{equation}
 \ind(m_R^{\mathrm{AIII}})=-1\in\mathbb Z,\qquad
 \ind(m_R^{\mathrm D})=1\in\mathbb Z_2.
 \label{ex-chain-labels}
\end{equation}
We have $R_{\mathrm{stab}}=2$ for both chains.

\subsection{\texorpdfstring{A $D_{3d}$ model motivated by bismuth}{A D3d model motivated by bismuth}}\label{ex-bismuth}

\paragraph{Hamiltonian.}
We use two finite-range lattice regularizations $H_1,H_3$ of effective bismuth models~\cite{schindlerHigherorderTopologyBismuth2018,zhaoRevealingHigherOrderTopological2025}. They act on four internal states per site of a triangular lattice extended along the third direction, with common reference $\varepsilon=-I\otimes\sigma_z$. Their real-space hopping blocks are given in \cref{ex-bi-model,ex-bi-hoppings}. We also consider $H_1\oplus H_3$ with the direct-sum reference.

The primitive vectors of the triangular layers and their unit interlayer spacing are
\begin{equation}
 \begin{gathered}
 \bm{a}_1=(1,0,0)^T,\qquad
 \bm{a}_2=(1/2,\sqrt3/2,0)^T,\\
 \bm{a}_3=(0,0,1)^T.
 \end{gathered}
 \label{ex-bi-lattice}
\end{equation}
Each site carries $\mathbb C^4$. Set
\begin{equation}
 \Gamma_0=I\otimes\sigma_z,\qquad
 (\Gamma_1,\Gamma_2,\Gamma_3)
 =(\sigma_z\otimes\sigma_x,-I\otimes\sigma_y,\sigma_x\otimes\sigma_x).
 \label{ex-bi-matrices}
\end{equation}
The two Hamiltonians $H_1$ and $H_3$ have translation-invariant blocks, so $(H_j)_{\bm{n},\bm{n}+\bm{\delta}}$ depends only on $\bm{\delta}$. Their onsite and reverse-hopping blocks satisfy
\begin{equation}
 \begin{aligned}
 (H_j)_{\bm{n},\bm{n}}&=-3\Gamma_0,\\
 (H_j)_{\bm{n}+\bm{\delta},\bm{n}}&=(H_j)_{\bm{n},\bm{n}+\bm{\delta}}^*,
 \qquad j=1,3,
 \end{aligned}
 \label{ex-bi-model}
\end{equation}
with one displacement from each nonzero hopping pair listed in \cref{ex-bi-hoppings}. All blocks not specified by the table or these relations vanish. The reference is $\varepsilon=-\Gamma_0$. These are finite-range lattice regularizations motivated by the effective models in Refs.~\cite{schindlerHigherorderTopologyBismuth2018,zhaoRevealingHigherOrderTopological2025}.

\begin{table}[htbp]
 \centering
 \begin{tabular}{c c c}
 \toprule
 Integer displacement $\bm{\delta}$ & $(H_1)_{\bm{n},\bm{n}+\bm{\delta}}$ & $(H_3)_{\bm{n},\bm{n}+\bm{\delta}}$\\
 \midrule
 $\bm{e}_1$ & $\frac12\Gamma_0-\frac{2i}{3}\Gamma_1$ & $\frac12\Gamma_0+i\Gamma_1$\\[3pt]
 $\bm{e}_2$ & $\frac12\Gamma_0-\frac{i}{3}\Gamma_1-\frac{i\sqrt3}{3}\Gamma_2$ & $\frac12\Gamma_0-i\Gamma_1$\\[3pt]
 $\bm{e}_2-\bm{e}_1$ & $\frac12\Gamma_0+\frac{i}{3}\Gamma_1-\frac{i\sqrt3}{3}\Gamma_2$ & $\frac12\Gamma_0+i\Gamma_1$\\[3pt]
 $\bm{e}_3$ & $\Gamma_0-i\Gamma_3$ & $\Gamma_0-i\Gamma_3$\\[3pt]
 $\bm{e}_1+\bm{e}_2$ & $0$ & $\frac{i}{3\sqrt3}\Gamma_2$\\[3pt]
 $2\bm{e}_2-\bm{e}_1$ & $0$ & $-\frac{i}{3\sqrt3}\Gamma_2$\\[3pt]
 $\bm{e}_2-2\bm{e}_1$ & $0$ & $\frac{i}{3\sqrt3}\Gamma_2$\\
 \bottomrule
 \end{tabular}
 \caption{Real-space hoppings of the $D_{3d}$ models. The reverse hopping is the adjoint.}
 \label{ex-bi-hoppings}
\end{table}

The bulk gaps satisfy $\gamma(H_1)>1.6$ and $\gamma(H_3)>0.2$. %

\paragraph{Symmetries and symmetry algebra.}
The symmetry group is $G=D_{3d}\times\mathbb Z_2^{T}$. The point group is generated by a sixfold improper rotation $s_6$ and a mirror $m_x$, and time reversal satisfies $U_T^2=-I$. All generators commute with $H_j$, so $h=0$; only $T$ is antilinear. The spinful projective relations, internal operators, and spatial actions are specified in \cref{ex-bi-site-actions,ex-bi-symmetry,ex-bi-relations}. They define
\begin{equation}
 \symalg=\mathbb C[G,\omega,s],\qquad
 \symalg^{V'}\cong
 \clifford^{3,3}\oplus\clifford^{3,3}\oplus M_2(\clifford^{3,3}).
 \label{ex-bi-algebra}
\end{equation}
The decomposition is obtained from the full symmetry algebra, including the action of $D_{3d}$ on $V'$.

\paragraph{Classification.}
Each factor in \cref{ex-bi-algebra} contributes an integer, giving the classification group $\mathbb Z^3$. For the energy scale fixed by \cref{ex-bi-model}, the labels at $R=8$ and their stability onsets are given in \cref{ex-bi-labels}.
\begin{table}[htbp]
 \centering
 \begin{tabular}{c c c}
 \toprule
 Hamiltonian & $\ind(m_8)$ & $R_{\mathrm{stab}}$\\
 \midrule
 $H_1$ & $(0,1,0)$ & $2$\\
 $H_3$ & $(0,-1,1)$ & $3$\\
 $H_1\oplus H_3$ & $(0,0,1)$ & $3$\\
 \bottomrule
 \end{tabular}
 \caption{Classification results for the bismuth models.}
 \label{ex-bi-labels}
\end{table}
All three labels are $(0,0,0)$ at $R=1$. At $R=2$, $H_3$ still has label $(0,0,0)$, while both $H_1$ and $H_1\oplus H_3$ have label $(0,1,0)$. The labels in \cref{ex-bi-labels} are obtained at every integer radius from the listed $R_{\mathrm{stab}}$ through $8$. The direct-sum Hamiltonian can be classified separately, and its label equals the sum of those of $H_1$ and $H_3$.

\subsection{Nonperiodic three-dimensional models}\label{ex-cubic}

\paragraph{Hamiltonian.}
We consider the two nearest-neighbor Hamiltonians in \cref{ex-cubic-hopping}, with reference $\varepsilon=I\otimes\sigma_z$. Each has four internal states per site and a nonperiodic onsite field that breaks the translation symmetry.

Both models use the unit cubic lattice with $\bm{a}_j=\bm{e}_j$ and $\bm{x}_{\bm{n}}=\bm{n}\in\mathbb Z^3$. Each site carries $\mathbb C^4$. Set
\begin{equation}
 \Gamma_0=I\otimes\sigma_z,\qquad
 (\Gamma_1,\Gamma_2,\Gamma_3)
 =(\sigma_x\otimes\sigma_x,\sigma_y\otimes\sigma_x,\sigma_z\otimes\sigma_x).
 \label{ex-cubic-matrices}
\end{equation}
The nonzero hopping blocks are
\begin{align}
 H_{\bm{n},\bm{n}}&=\bigl(-2+\tfrac{\nu(\bm{n})}{10}\bigr)\Gamma_0,\nonumber\\
 H_{\bm{n},\bm{n}+\bm{e}_j}&=\tfrac12(-\Gamma_0-i\Gamma_j),\nonumber\\
 H_{\bm{n}+\bm{e}_j,\bm{n}}&=H_{\bm{n},\bm{n}+\bm{e}_j}^*,\qquad j=1,2,3.
 \label{ex-cubic-hopping}
\end{align}
The reference is $\varepsilon=\Gamma_0$, and the nonperiodic onsite field is
\begin{equation}
 \nu(\bm{n})=\frac{(-1)^{b(|n_1|)+b(|n_2|)+b(|n_3|)}}{1+|\bm{n}|^2},
 \label{ex-disorder}
\end{equation}
where $|\bm{n}|^2=n_1^2+n_2^2+n_3^2$ and $b(q)$ counts the ones in the binary expansion of a nonnegative integer $q$. For example, $5=(101)_2$ has two ones, so $b(5)=2$. Since $\nu(\bm{0})=1$ and $\nu(\bm{n})$ tends to zero along every nonzero lattice direction, this field has no nonzero translation period.

The clean Hamiltonian has bulk gap $1$, and the onsite perturbation has norm $1/10$, so both models satisfy $\gamma(H)\ge 9/10$.

\paragraph{Symmetries and symmetry algebra.}
The AII model has onsite time reversal $T$ with $U_T^2=-I$. The DIII model also has onsite particle-hole symmetry $C$ with $U_C^2=I$ and $U_TU_C=U_CU_T$. Their matrices are given in \cref{ex-cubic-symmetry}. The gradings are $(s,h)(T)=(1,0)$ and $(s,h)(C)=(1,1)$. No point-group constraint is imposed on either model. The physical symmetry algebras are
\begin{equation}
 \symalg(\mathrm{AII})\cong\mathbb H,\qquad
 \symalg(\mathrm{DIII})\cong\clifford^{0,3}.
 \label{ex-cubic-source-algebras}
\end{equation}
Since the antiunitary symmetries fix sites, each acts as $-I$ on the three-dimensional $V'$. The resulting target algebras are listed in \cref{ex-three-labels}.

\paragraph{Classification.}
The symmetry algebras give $\mathbb Z_2$ and $\mathbb Z$ for AII and DIII, respectively. The labels at $R=8$ and their stability onsets are given in \cref{ex-three-labels}.
\begin{table}[htbp]
 \centering
 \begin{tabular}{c c c c c}
 \toprule
 Class & $\symalg^{V'}$ & Classification group & $\ind(m_8)$ & $R_{\mathrm{stab}}$\\
 \midrule
 AII & $M_4(\clifford^{0,1})$ & $\mathbb Z_2$ & $1$ & $3$\\
 DIII & $\clifford^{3,3}$ & $\mathbb Z$ & $-1$ & $3$\\
 \bottomrule
 \end{tabular}
 \caption{Classification results for the nonperiodic three-dimensional models.}
 \label{ex-three-labels}
\end{table}
The AII and DIII inputs both have label $0$ at $R=1,2$; their respective labels $1$ and $-1$ are obtained at every integer radius $R=3,\ldots,8$.

\subsection{A magnetic spin-point superconductor with nontrivial pairing}\label{ex-spin-pairing}

\paragraph{Hamiltonian.}
We construct a three-dimensional superconductor realizing the magnetic spin point group $8.87$ in the Bilbao SPGENPOS catalog~\cite{Elcoro2026SpinPointGroups}, corresponding to Litvin No.~87 in Ref.~\cite[Table~7]{schiffCrystallographicSpinPoint2025}. The model has two orbital states, two spin states, and their Nambu partners at each site of the unit-spaced cubic lattice, with nearest-neighbor pairing in all three directions. The real-space Hamiltonian and reference are specified in \cref{ex-spin-matrices,ex-spin-hopping}.

Here $\bm{a}_j=\bm{e}_j$, so $\bm{x}_{\bm{n}}=\bm{n}\in\mathbb Z^3$. Each site carries $\mathbb C^2\otimes\mathbb C^2\otimes\mathbb C^2$, in the tensor order Nambu, orbital, spin. Define
\begin{align}
 \Gamma_0&=\sigma_z\otimes I\otimes I,&
 \Gamma_1&=\sigma_x\otimes\sigma_x\otimes\sigma_x,\nonumber\\
 \Gamma_2&=-\sigma_y\otimes\sigma_z\otimes\sigma_z,&
 \Gamma_3&=\sigma_x\otimes I\otimes\sigma_z.
 \label{ex-spin-matrices}
\end{align}
The nonzero real-space blocks and the reference are
\begin{align}
 H_{\bm{n},\bm{n}}&=2\Gamma_0,\nonumber\\
 H_{\bm{n},\bm{n}+\bm{e}_j}&=\frac12(-\Gamma_0-i\Gamma_j),\nonumber\\
 H_{\bm{n}+\bm{e}_j,\bm{n}}&=H_{\bm{n},\bm{n}+\bm{e}_j}^*,\qquad j=1,2,3,\nonumber\\
 \varepsilon&=\Gamma_0.
 \label{ex-spin-hopping}
\end{align}
All unlisted blocks vanish.

The bulk gap satisfies $\gamma(H)\ge1$.

\paragraph{Symmetries and symmetry algebra.}
In the catalog setting, the three spatial mirrors $m_x,m_y',m_z'$ generate $G_0\cong(\mathbb Z_2)^3$. Their accompanying spin operations are a twofold rotation about $z$, spin reversal, and reflection in the spin plane normal to $x$, respectively. The primes denote antiunitary symmetries. The superconductor pairing character is $(\chi(m_x),\chi(m_y'),\chi(m_z'))=(-1,1,-1)$. All three mirrors preserve the Hamiltonian. Adding the intrinsic particle-hole symmetry $C$, with $(s,h)(C)=(1,1)$, gives $G=G_0\times\mathbb Z_2^C$. The spin actions and the superconductor pairing both enter the projective multiplier $\omega$, as shown in \cref{ex-spin-actions,ex-spin-relations}.

The symmetry algebra and its three-dimensional target are
\begin{equation}
 \symalg=\mathbb C[G,\omega,s],\qquad
 \symalg^{V'}\cong
 M_8(\clifford^{0,1})\oplus M_8(\clifford^{0,1}).
 \label{ex-spin-algebra}
\end{equation}

\paragraph{Classification.}
The two simple factors in \cref{ex-spin-algebra} give the classification group $\mathbb Z_2\oplus\mathbb Z_2$. The numerical computation gives
\begin{equation}
 \ind(m_R)=(1,0)\in\mathbb Z_2\oplus\mathbb Z_2,
 \qquad R=3,\ldots,8.
 \label{ex-spin-label}
\end{equation}
At $R=1,2$, the label is $(0,0)$, so $R_{\mathrm{stab}}=3$ on the integer-radius scan.

\section{Conclusion}\label{sec:conclusion}

We have presented a construction that connects local gapped free-fermion Hamiltonians in real space to finite-size zero-dimensional systems, without assuming translation symmetry. The general symmetry data $(G,\omega,s,h,V)$ provide a common description of spatial actions, antiunitary operations, antisymmetries, and projective phases. Through the symmetry algebra, the same construction accommodates magnetic point groups, finite spin point groups, and nontrivial superconducting pairing representations. The sphere construction and its discretization also give a classifying-space interpretation of the spectral localizer. For a gapped finite-size zero-dimensional system, the numerical calculation determines its topological class using FreeFermionicSPT~\cite{ttiiddeeTtiiddeeFreeFermionicSPT2026}.

The equivalence of the real-space and zero-dimensional classifying spaces remains the conjecture stated in Conjecture~\ref{conj-real-space}. Establishing its precise topological formulation and proving the comparison equivalences would provide a foundation for interpreting the resulting indices as a complete real-space classification.

\begin{acknowledgments}
The authors thank Chen Fang for insightful discussions. Y.Q. acknowledges support from National Key R\&D Program of China (Grant No. 2022YFA1403402), from the National Natural Science Foundation of China (Grant No. 12174068), and from the Science and Technology Commission of Shanghai Municipality (Grants No. 24LZ1400100 and No. 23JC1400600).

We used generative AI tools to help draft portions of the manuscript, write parts of the computational code, and construct some of the model examples in Sec.~\ref{ex-section}.
\end{acknowledgments}

\appendix
\section{Model symmetries}
\label{app:example-models}

We specify the symmetry operators used in \cref{ex-bismuth,ex-cubic,ex-spin-pairing}, with the notation of \cref{ex-section}. Symmetry operators on each site are $U_g$, including complex conjugation when $s(g)=1$. The physical operators combine these onsite matrices with the site permutations. All models in this appendix have $d=3$.

\subsection{\texorpdfstring{The $D_{3d}$ models}{The D3d models}}
\label{ex-bi-input-appendix}

Let $s_6$ be a sixfold improper rotation and $m_x$ the mirror reversing the first Cartesian coordinate. Their site actions and that of time reversal are
\begin{equation}
 \begin{aligned}
 s_6:\bm{n}&\mapsto(n_1+n_2,-n_1,-n_3),\\
 m_x:\bm{n}&\mapsto(-n_1-n_2,n_2,n_3),\\
 T:\bm{n}&\mapsto \bm{n}.
 \end{aligned}
 \label{ex-bi-site-actions}
\end{equation}
The onsite symmetry actions are given by
\begin{align}
 U_{s_6}^{(1)}
 &=\operatorname{diag}(e^{-2\pi i/3},-e^{2\pi i/3},
                       e^{2\pi i/3},-e^{-2\pi i/3}),\nonumber\\
 U_{s_6}^{(3)}&=\Gamma_0,\nonumber\\
 U_{m_x}&=i\sigma_x\otimes I,
 \qquad U_T=(\sigma_y\otimes I)\mathsf K.
 \label{ex-bi-symmetry}
\end{align}
The physical operators combine these onsite matrices with the site permutations in \cref{ex-bi-site-actions}, as in \cref{eq-hilbert-action}. Their projective relations are
\begin{equation}
 \begin{gathered}
 s_6^6=1,\qquad m_x^2=T^2=-1,\\
 m_xs_6m_x^{-1}=s_6^{-1},\qquad
 [T,s_6]=[T,m_x]=0.
 \end{gathered}
 \label{ex-bi-relations}
\end{equation}
Here $s_6$ and $m_x$ are linear, $T$ is antilinear, and all three commute with $H_j$. These relations fix the multiplier used in $\symalg=\mathbb C[D_{3d}\times\mathbb Z_2^{T},\omega,s]$. In $V'$, time reversal acts as $-I$, while the unitary spatial actions are unchanged. The direct-sum model uses direct sums of the hoppings, symmetry operators, and reference.

\subsection{Nonperiodic three-dimensional models}
\label{ex-cubic-input-appendix}

The antiunitary operators are onsite:
\begin{equation}
 U_T=(i\sigma_y\otimes I)\mathsf K,\qquad
 U_C=(\sigma_y\otimes\sigma_y)\mathsf K.
 \label{ex-cubic-symmetry}
\end{equation}
AII imposes $T$; DIII imposes both $T$ and $C$. They satisfy $T^2=-1$, $C^2=1$, and $TC=CT$, with $T HT^{-1}=H$ and $C HC^{-1}=-H$. No spatial point group is imposed.

\subsection{The magnetic spin-point superconductor}
\label{ex-spin-input-appendix}

\paragraph{Magnetic spin point group.}
A magnetic spin point group consists of symmetries that pair a lattice point operation with an orthogonal transformation of the three spin components. Such independent spatial and spin actions arise when spin--orbit coupling is negligible~\cite{liuSpinGroupSymmetryMagnetic2022,schiffCrystallographicSpinPoint2025}.

The input consists of an abstract group $G_0$ and two orthogonal representations, $g\mapsto\rho_V(g)\in O(V)$ on real space and $g\mapsto O_g\in O(3)$ on spin. It suffices to specify the pair $(\rho_V(g),O_g)$ for each generator. Group $8.87$ in SPGENPOS~\cite[Supplemental Material, Sec.~S1.1]{Elcoro2026SpinPointGroups} is Litvin No.~87 in Ref.~\cite[Table~7]{schiffCrystallographicSpinPoint2025}. Its symbol in the catalog axes is
\begin{equation}
 {}^{2_z}m\,{}^{\overline{1}}m\,{}^{m_x}m.
 \label{ex-spin-catalog-setting}
\end{equation}
The three mirrors have spatial normals $x,y,z$, in that order. Their superscripts give the accompanying spin operations, respectively a rotation by $\pi$ about $z$, reversal of all three spin components, and reflection in the plane normal to $x$. In these axes $G_0=\langle m_x,m_y',m_z'\rangle\cong(\mathbb Z_2)^3$, with generator pairs
\begin{equation}
 \begin{array}{c|ccc}
 g&m_x&m_y'&m_z'\\ \hline
 \rho_V(g)&\operatorname{diag}(-1,1,1)&\operatorname{diag}(1,-1,1)&\operatorname{diag}(1,1,-1)\\[2pt]
 O_g&\operatorname{diag}(-1,-1,1)&-I_3&\operatorname{diag}(-1,1,1)
 \end{array}
 \label{ex-spin-parent-actions}
\end{equation}

The antiunitary grading is determined by $(-1)^{s(g)}=\det O_g$. Thus $m_x$ is unitary ($s=0$), and the primed generators $m_y'$ and $m_z'$ are antiunitary ($s=1$).

For a spin-$\tfrac12$ electron, choose a lift $\widetilde O_g$ of each $O_g$ under the double cover $\mathrm{Pin}_+(3)\to O(3)$, with $\widetilde O_1=1$ and
\begin{equation}
 \omega(g,k)=\widetilde O_g\widetilde O_k\widetilde O_{gk}^{-1}\in\{\pm1\},\qquad g,k\in G_0.
 \label{ex-spin-lift}
\end{equation}
For an ordinary magnetic point group with real-space action $\rho_V(g)$, the spin action is $O_g=(-1)^{s(g)}(\det \rho_V(g))\rho_V(g)$ and the projective phase is $\omega(g,k)=(-1)^{s(g)s(k)}p_gp_kp_{gk}^{-1}$, where $p_g$ lifts $\rho_V(g)$ to $\mathrm{Pin}_-(3)$ and $p_1=1$. These two constructions yield equivalent $\omega$ for an ordinary magnetic point group.

\paragraph{Superconducting pairing.}
Under $g\in G_0$ the upper-right Nambu block $\Delta_{\bm{n},\bm{n}'}$ transforms to $\chi(g)\Delta_{g\bm{n},g\bm{n}'}$. The mixed product with particle-hole symmetry $C$ is given in Ref.~\cite[Eq.~(84)]{shiozakiClassificationSurfaceStates2022} for magnetic symmetries. With $U_{gC}=U_gU_C$ we obtain
\begin{gather}
 U_CU_g=\overline{\chi(g)}\,U_{gC},\qquad
 \omega(C,g)=\overline{\chi(g)},\quad \omega(g,C)=1,\label{ex-spin-relations}\\
 \bigl(\chi(m_x),\chi(m_y'),\chi(m_z')\bigr)=(-1,1,-1).
 \label{ex-spin-pairing-data}
\end{gather}

The symmetry operators for $G=G_0\times\mathbb Z_2^C$ are
\begin{align}
 U_{m_x}&=iI\otimes I\otimes\sigma_z,&
 U_{m_y'}&=(iI\otimes I\otimes\sigma_y)\mathsf K,\nonumber\\
 U_{m_z'}&=(-iI\otimes I\otimes\sigma_z)\mathsf K,&
 U_C&=(\sigma_x\otimes I\otimes I)\mathsf K.
 \label{ex-spin-actions}
\end{align}
In each matrix, the middle tensor factor is the identity on the two orbital states. Combined with their spatial actions, the three mirrors preserve $H$ and $\varepsilon$, while $C$ reverses both.

\bibliography{zotero}
\end{document}